\documentclass[submission,copyright,creativecommons]{eptcs}
\providecommand{\event}{FESCA 2016} 

\usepackage[utf8]{inputenc} 
\usepackage{url} 
\usepackage{xcolor} 
\usepackage{soul} 
\usepackage[normalem]{ulem}
\usepackage{graphicx}
\usepackage{paralist}

\usepackage{listings}
\usepackage{framed}

\definecolor{irmexchangecolor}{HTML}{AAE8FF}
\definecolor{irmprocesscolor}{HTML}{CCFF99}
\definecolor{irmassumptioncolor}{HTML}{FFFF6C}
\definecolor{irmabstractcolor}{HTML}{E5E5E5}

\newcommand{\irmassumption}[1]{ {\sethlcolor{irmassumptioncolor} \hl{#1}} }
\newcommand{\irmprocess}[1]{ {\sethlcolor{irmprocesscolor} \hl{#1}} }
\newcommand{\irmexchange}[1]{ {\sethlcolor{irmexchangecolor} \hl{#1}} }
\newcommand{\irmabstract}[1]{ {\sethlcolor{irmabstractcolor} \hl{#1}} }

\newcommand{\irminvariantid}[1]{{\textcolor{violet}{[#1]}}}

\def\ulcar#1{\color{red}\underline{\underline{{\color{black}#1}}}\color{black}}
\def\ulcaratt#1{\color{red}\underline{{\color{black}#1}}\color{black}}

\def\ulpark#1{\color{blue}\underline{\underline{{\color{black}#1}}}\color{black}}
\def\ulparkatt#1{\color{blue}\underline{{\color{black}#1}}\color{black}}

\title{Towards an Automated Requirements-driven Development of Smart Cyber-Physical Systems \\ \Large Position paper}

\author{Jiri Vinarek \qquad\qquad\qquad Petr Hnetynka
\institute{Charles University in Prague, Faculty of Mathematics and Physics,\\ 
Department of Distributed and Dependable Systems,\\ 
Malostranske namesti 25, Prague, Czech Republic}
\email{vinarek@d3s.mff.cuni.cz \qquad\qquad\qquad hnetynka@d3s.mff.cuni.cz}
}
\def\titlerunning{Towards an Automated Requirements-driven Development of sCPS}
\def\authorrunning{J. Vinarek, P. Hnetynka}
\begin{document}
\maketitle

\begin{abstract}
The Invariant Refinement Method for Self Adaptation (IRM-SA) is a design method targeting development of smart Cyber-Physical Systems (sCPS).
It allows for a systematic translation of the system requirements into the system architecture expressed as an ensemble-based component system (EBCS).
However, since the requirements are captured using natural language, there exists the danger of their misinterpretation due to natural language requirements’ ambiguity, which could eventually lead to design errors.
Thus, automation and validation of the design process is desirable.
In this paper, we (i) analyze the translation process of natural language requirements into the IRM-SA model, (ii) identify individual steps that can be automated and/or validated using natural language processing techniques, and (iii) propose suitable methods.

\end{abstract}

\section{Introduction}







Smart Cyber-Physical Systems (sCPS) are complex distributed decentralized systems of cooperating mobile and stationary devices closely interacting with the physical environment.
Examples of sCPS include systems like smart home, smart traffic management, etc.

Designing and developing such a system is a quite complex task with many challenges. 
Mobility and distribution bring the high level of dynamism to the system, which has to be aware of changes in its environment.
Openness and open-endness are other challenging issues resulting in needs that the designed system has to be able to tackle unanticipated changes and participants unknown at design time.

The traditional software design and development techniques have been shown unsuitable for such systems and novel approaches~\cite{Holzl_Ensembles_2008,Morin_Taming_2015,Ruchkin_Hybrid_2015} have been proposed to tackle with the challenges.
One of these promising approaches is Ensemble-Based Component Systems (EBCS)~\cite{Bures_Deeco_2013}.
Using EBCS, the system is modeled and developed as a set of \emph{ensembles}, i.e., dynamic cooperation groups of software components.
Components are specified by their \emph{knowledge} (i.e., component's attributes) and by a set of \emph{processes} manipulating the knowledge.


The Invariant Refinement Method for Self Adaptation (IRM-SA) \cite{BuresIrm15} is a design method targeting development of sCPS using EBCS.
IRM-SA allows for a systematic translation of the system requirements written in natural language into the system architecture expressed as components and ensembles.
Using IRM-SA, a designer gradually refines the initial requirements and iteratively builds a model that consists of so-called \textit{invariants}. 
Invariants are then hierarchically decomposed and at the lowest level, they directly correspond to an implementation (in the DEECo component model \cite{Bures_Deeco_2013}, with which IRM-SA is currently tied).

To speedup and ease the design process with IRM-SA, the guide \footnote{\url{http://svn.pst.ifi.lmu.de/ascens/guide/irm/}} and graphical editor \footnote{\url{https://github.com/d3scomp/IRM-SA}} have been created. 
The editor allows for editing of the constraints and performing several basic validations of the designed IRM-SA model.
Additionally, skeletons of the implementation can be generated directly from the designed model.
Even though the guide and editor exist, the whole process, i.e., translation of requirements into the IRM-SA model, is manual and it can be time-consuming and laborious.
Additionally, as the requirements are expressed as a text in natural language, there is a danger of ambiguity and misinterpretation of them, which can result in a suboptimal design.
Even more, designers can unintentionally miss important requirements.
%

The goal of this paper is to analyze the IRM-SA design process and identify particular steps, which can be, fully or at least partially, automated with the help of natural language processing tools.
To achieve the goal, we use our experience gained with automated processing of textual use-cases, their verification and transformation into an implementation (\cite{simko2014foam,Simko2010,vinarek_recovering_2014}).


The paper is structured as follows:
Section~\ref{sec:irm-sa-explained} explains the IRM-SA method and its inputs and outputs.
Section~\ref{sec:automation} discusses steps of the IRM-SA method from the perspective of their automation and proposes solutions for them.
Finally, Section~\ref{sec:related} discusses related work while Section~\ref{sec:conclusion} concludes the paper.

\section{IRM-SA explained by example}
\label{sec:irm-sa-explained}

In this section, we briefly describe the IRM-SA method and its usage on an example (for a detailed description, please see the IRM-SA guide).
The experiment described in \cite{IliasTr15} proved that usage of IRM-SA represents a significant help in EBCS design and development.
Participants of this experiment designed using IRM-SA an EBCS architecture with less errors than participants using another design method. 
Even though, the resulting architectures were not completely without errors, especially thank to different understanding of the input requirements provided as a text in natural language and thus, there is still space for improvements. 

This is even more important, as one of the outputs of the IRM-SA method~--~the IRM-SA model~--~can be used not only at design time, but also during development and maintenance of the developed system.
In particular, the IRM-SA model allows for traceability between purpose of each invariant (requirement) and its realization and therefore it is ideal for documentation and maintenance.
Plus, as stated in~\cite{Larman2004}, it is a mistake to understand requirements specifications as final and unchangeable and thus keeping up-to-date traceability links to requirements is quite important.

Additionally, the IRM-SA model is in the DEECo implementation employed for controlling self-adaptation of the system, i.e., the model captures multiple alternatives of the system architecture and the appropriate one is chosen based on actual situation.
 

To sum up, the IRM-SA model is one of the key artifacts of the developed system and its correctness is essential.
Therefore, designers/developers would benefit from a tool which not only allows for easy creation of the model (the currently available editor allows for this) but also which would be able to (semi)automatically parse the textual requirements, generate parts of the model from the requirements, and validate individual actions performed by the designer/developer. 
To provide such a tool, natural language processing methods and tools have to be incorporated in the process.

\subsection{IRM-SA method and model}

The IRM-SA design method is an iterative top-down design approach.
A designer has to perform the following steps in order to built the IRM-SA model from the requirement specification:
\begin{enumerate}
\item Find the top-level goals of the system and specify the top-level (abstract) invariants.
\item Find the components of the system (and their fields) by asking ``which knowledge does each invariant involve and where is this knowledge obtained from?''
\item Decompose each invariant by asking “how can this invariant be satisfied?”
\item Separate the concerns of the abstract invariants into sub-invariants that correspond to (abstract) activities that can be done in isolation.
\item Compose invariants together by asking “why do I need to satisfy these invariants?”
\item In case of situation-specific requirements, try first to accurately capture the condition of being in one situation or another. Use the assumptions to do that. Then use OR decomposition to specify which invariants to satisfy in each situation.
\end{enumerate}

\begin{figure}[ht]
	\centering
	\includegraphics[width=\linewidth]{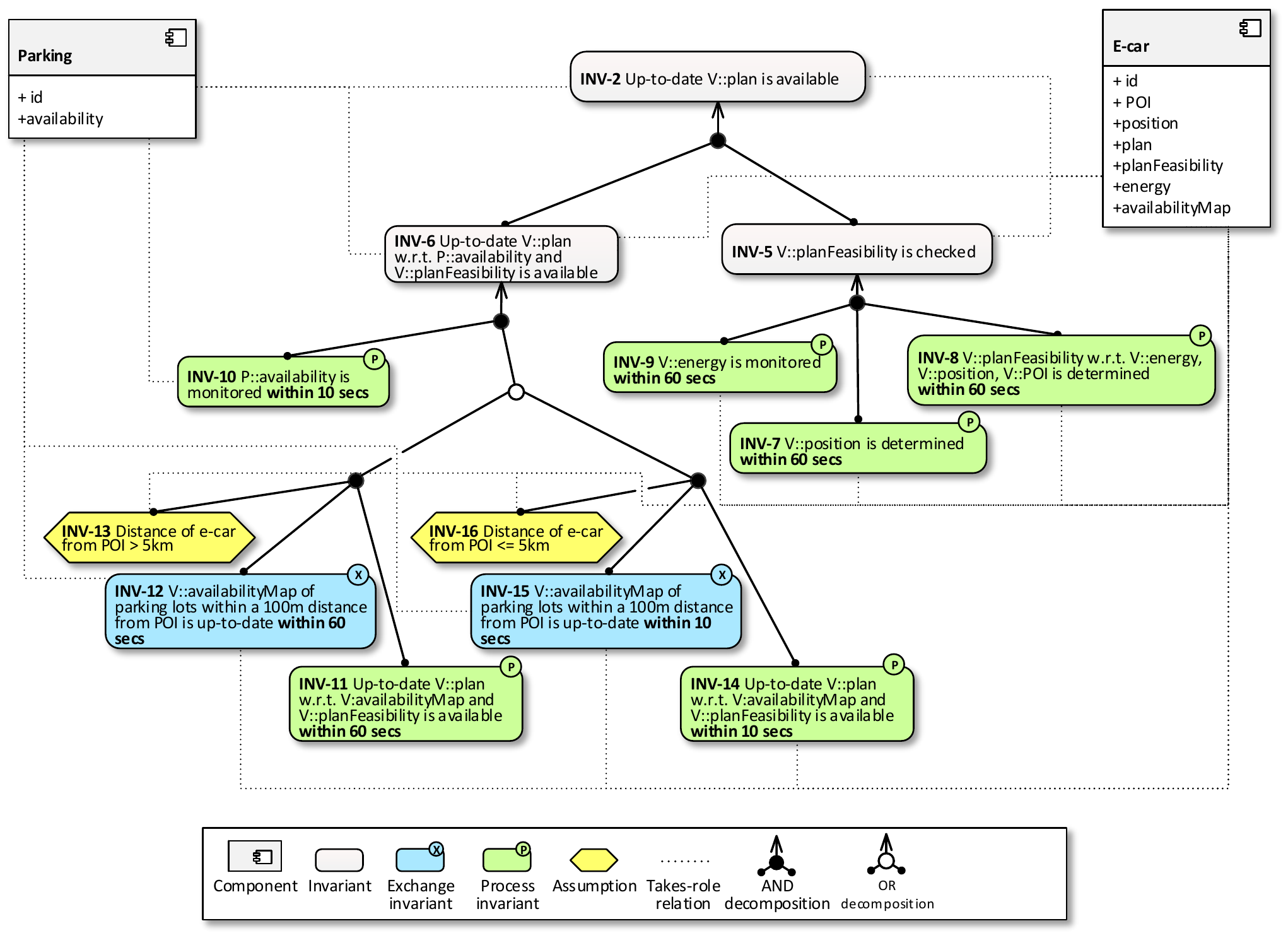}
	\caption[e-cars-irm-model]{IRM-SA model for e-cars system}
	\label{fig:irm-e-cars-model}
\end{figure}

\begin{figure}[h]
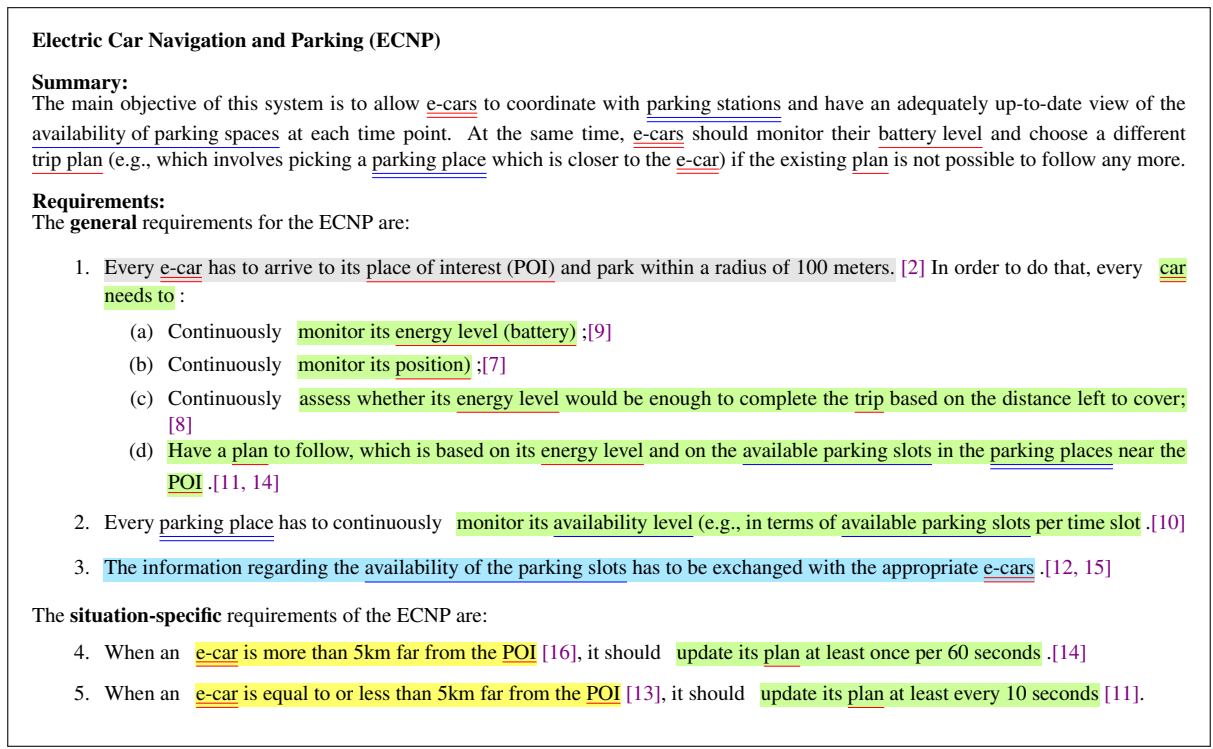


\begin{framed}
\fontsize{8}{8}\selectfont

\textbf{Electric Car Navigation and Parking (ECNP)}\\

\textbf{Summary:}\\
The main objective of this system is to allow {\ulcar{e-cars}} to coordinate with {\ulpark{parking stations}} and have an adequately up-to-date view of the {\ulparkatt{availability of parking spaces}} at each time point. At the same time, {\ulcar{e-cars}} should monitor their {\ulcaratt{battery level}} and choose a different  {\ulcaratt{trip plan}} (e.g., which involves picking a {\ulpark{parking place}} which is closer to the {\ulcar{e-car}}) if the existing {\ulcaratt{plan}} is not possible to follow any more.\\

\textbf{Requirements:}\\
The \textbf{general} requirements for the ECNP are:

\begin{enumerate}
\item \irmabstract{Every {\ulcar{e-car}}  has to arrive to its {\ulcaratt{place of interest (POI)}} and park within a radius of 100 meters.}\irminvariantid{2} In order to do that, every \irmprocess{{\ulcar{car}} needs to}:
\begin{enumerate}
\item Continuously \irmprocess{monitor its {\ulcaratt{energy level (battery)}}};\irminvariantid{9}
\item Continuously \irmprocess{monitor its {\ulcaratt{position)}}};\irminvariantid{7}
\item Continuously \irmprocess{assess whether its {\ulcaratt{energy level}} would be enough to complete the {\ulcaratt{trip}} based on the distance left to cover;}\irminvariantid{8}
\item \irmprocess{Have a {\ulcaratt{plan}} to follow, which is based on its {\ulcaratt{energy level}} and on the {\ulparkatt{available parking slots}} in the {\ulpark{parking places}} near the {\ulcaratt{POI}}}.\irminvariantid{11, 14}
\end{enumerate}
\item Every {\ulpark{parking place}} has to continuously \irmprocess{monitor its {\ulparkatt{availability level}} (e.g., in terms of {\ulparkatt{available parking slots}} per time slot}.\irminvariantid{10}
\item \irmexchange{The information regarding the {\ulparkatt{availability of the parking slots}} has to be exchanged with the appropriate {\ulcar{e-cars}}}.\irminvariantid{12, 15}
\end{enumerate}

\noindent
The \textbf{situation-specific} requirements of the ECNP are:
\begin{enumerate}
\setcounter{enumi}{3}
\item When an \irmassumption{{\ulcar{e-car}} is more than 5km far from the {\ulcaratt{POI}}}\irminvariantid{16}, it should \irmprocess{update its {\ulcaratt{plan}} at least once per 60 seconds}.\irminvariantid{14}
\item When an \irmassumption{{\ulcar{e-car}} is equal to or less than 5km far from the {\ulcaratt{POI}}}\irminvariantid{13}, it should \irmprocess{update its {\ulcaratt{plan}} at least every 10 seconds}\irminvariantid{11}.
\end{enumerate}
\end{framed}
\caption[e-cars-requirements]{e-cars system requirements}
\label{fig:e-cars-requirements}
\end{figure}

Figure~\ref{fig:irm-e-cars-model} shows the IRM-SA model created for the electric car (e-car) navigation and parking system example; its requirement specification is in Figure~\ref{fig:e-cars-requirements}. 
Both the specification and model are overtaken from the IRM-SA guide (and they were also employed in the experiment mentioned above).
The model was created manually with the help of the IRM-SA model editor.

The highlighting and underlining in the requirements specification is included solely for the purpose of this paper and would not appear in an actual specification.
Meaning of the highlighting/underlining is as follows:
\begin{itemize}
\item Underlined text refers to components and their attributes (double line for components, single line for attributes). Red color is used for the ``E-car'' component, blue one for the ``Parking'' component)
\item Highlighted text is related to a particular invariant; colors correspond with colors in figure \ref{fig:irm-e-cars-model}.
\item Purple numbers in the requirements specification refer to corresponding invariants in the model.
\end{itemize} 

\section{Automation of IRM-SA}
\label{sec:automation}

In this section, we analyze the IRM-SA method from the perspective of its automation. 
We identify individual steps that can be automated using natural language processing techniques, and propose suitable methods.

The individual goals of such an automation are:
\begin{inparaenum}[(i)]
\item (semi)automatically generate invariants in the IRM-SA model from the requirements document (and thus make synchronization and traceability between the requirements and IRM-SA model more robust and faster to obtain),
\item (semi)automatically validate the resulting IRM-SA model.
\end{inparaenum}

\subsection{Component identification}
\label{sec:component_identification}

%

Components in EBCS design represent ``smart'' entities of the system.
In our example, there are two types of components~--~\texttt{E-Car} and \texttt{Parking}.
Both components and also their attributes are several times mentioned in the \textit{requirements} as well as in the \textit{summary}.
Based on our experience with derivation of the domain model from textual specification \cite{SimkoTr13}, it seems possible to obtain a list of potential components in an automated fashion.
Also, a similar approach is employed in \cite{DrechslerSW12}, where authors retrieve UML class models from test cases.

\begin{figure}[ht]
	\centering
	\includegraphics[width=\linewidth]{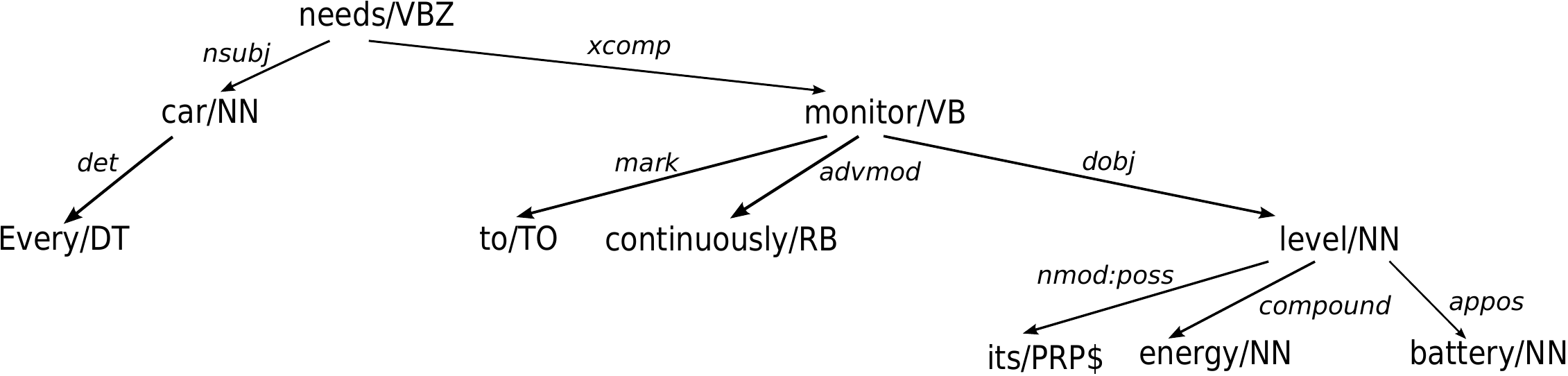}
	\caption[stanford-deps]{Dependencies and POS tags obtained from Stanford dependency parser}
	\label{fig:stanford-dependencies-example}
\end{figure}

Both names of the components and their attributes are in the requirement texts almost always represented as noun phrases, which in simple sentences appear as \textit{subjects} or \textit{objects} (either direct or indirect).
To parse a sentence and identify its elements, the Stanford CoreNLP toolkit \cite{manning-EtAl:2014:P14-5} is an ideal tool. 
For example, with the usage of the Stanford dependency parser on the sentence \textit{"Every car needs to continuously monitor its energy level (battery)."}, we get the dependency graph showed in Figure~\ref{fig:stanford-dependencies-example}.
The parser returns a part-of-speech (POS) tag for each word (e.g., \textit{VB} for verb in base form, \textit{NN} for singular noun) and dependency relations between the words (e.g., \textit{nsubj} for nominal subject or \textit{dobj} for direct object).
The resulting list of potential names of components and attributes contains \texttt{car} (nominal subject) and \texttt{energy level} (direct object).
Unfortunately, this might not be sufficient (not all of the subjects/objects are components/attributes).
A possible way to overcome this issue is to employ statistical classification techniques to learn the patterns from training data. Similarly, we have employed these techniques in~\cite{vinarek_recovering_2014}.

\subsection{Component disambiguation}
\label{sec:component_disambiguation}

Another issue with the list of candidates for component/attribute names is that it may be ambiguous. 
Multiple noun phrases may refer to the same component, e.g., in our e-cars system specification the words ``e-car'' and ``car'' refer to the same component. 
Similarly,  ``parking station'' and ``parking place'' or ``energy level'' and ``battery''. 
These ambiguities can be a sign of a poorly written specification and their replacement with the same word or phrase is advisable.
On the other hand, such a situation may happen (especially if the specification is prepared by multiple authors and/or it evolves over time) and even more, in some cases, the use of different words for the same entity may be intentional, e.g., in a case of abbreviations and noun phrase shortenings (``place of interest'' and ``POI'' or ``trip plan'' and ``plan'', etc.).

A distinction of these cases is not always clear and we expect that user will have to be involved in a decision about these ambiguities.

For example, the requirement in Figure~\ref{fig:stanford-dependencies-example} is written in a way which suggest that the phrase ``energy level'' and ``battery'' can be used interchangeably.
This relation can be deduced automatically, as the word ``battery'' has been marked as appositional modifier (\textit{appos}) of the word ``level''.
Another option for disambiguation might be employment of \textit{string distance metrics}~\cite{Cohen03acomparison}\footnote{an implementation available at \url{http://secondstring.sourceforge.net/}} to identify corresponding entities (e.g., ``car'' and ``e-car'').


\subsection{Invariant type identification}
\label{sec:invariant_type_identification}

During applying the IRM-SA method, sentences in the \textit{Requirements} section of the specification are rather directly translated into invariants of the IRM-SA model.
However, the issue is to identify whether the particular sentence relates to the \textit{abstract}, \textit{process}, \textit{exchange} or \textit{assumption} invariant.
It would be helpful, if the IRM-SA editor could automatically propose the invariant type.

\textit{Assumption} invariants should be included only in the \textit{situation-specific} section of the requirements specification and thus is easier to locate them (see the yellow highlighting in Figure~\ref{fig:e-cars-requirements} and the yellow invariants in Figure~\ref{fig:irm-e-cars-model}).
Additionally, the particular sentences express a condition, which is necessary to detect and extract.
To extract it, the dependency parser can be again utilized. 
To support it, tools for information extraction like Ollie\footnote{\url{https://github.com/knowitall/ollie}} or OpenIE\footnote{\url{https://github.com/knowitall/openie}} can also be used as they are able to detect enabling conditions.


To distinguish between \textit{process} and \textit{exchange} invariants, it is necessary to analyze the main verb of the sentence (see the blue and green highlighting in Figure~\ref{fig:e-cars-requirements} and the blue and green invariants in Figure~\ref{fig:irm-e-cars-model}). 
With verbs such as ``exchange'' or ``propagate'', there is a high chance that the sentence corresponds to exchange invariant, while verbs ``have'', ``monitor'', ``assess'', ``obtain'', ``acquire'' or ``determine'' usually denote a process invariant.
A direct solution would be a simple comparison of the particular verb with a predefined set of verbs but it would be rather limiting.
Instead, a suitable approach is to classify verbs according to their meaning, which is taken from WordNet\cite{Miller:1995:WLD:219717.219748}.
WordNet is a large lexical database of English nouns, verbs, adjectives and adverbs.
In the database, synonyms are grouped together forming so-called \textit{synsets}. 
The synsets are interlinked according to their relations, forming network of related words and concepts.
Multiple WordNet similarity measures were proposed and their implementation is available\footnote{\url{http://search.cpan.org/dist/WordNet-Similarity/}} and can be used for the \textit{process} and \textit{exchange} invariants identification.


Finally, sentences containing additional sub-requirements can be marked as \textit{abstract} invariants (the gray highlighting and the gray invariants in Figures~\ref{fig:irm-e-cars-model} and \ref{fig:e-cars-requirements}).


\subsection{Knowledge flow recognition}
\label{sec:knowledge_flow_recognition}

One of the key IRM-SA ideas is that each invariant (with the exception of assumption ones) represents a computation that produces \textit{output knowledge} given a particular \textit{input knowledge} such that the invariant is satisfied (as stated in~\cite{BuresIrm15}). 

For example, in the invariant deduced from the requirement 1(d) in Figure~\ref{fig:e-cars-requirements}, the \textit{energy level} and \textit{POI} of the \textit{vehicle} and the \textit{available parking slots} from the \textit{parking places} serve as the input parameters for computation of the vehicle's \textit{plan} (all possible parameters, i.e., component attributes, are already known as they were identified in the previous phases).
Schematically, it can be written 

\begin{center}
\texttt{V::energy, V::POI, P::availability -> V::plan}.
\end{center}
Such an abstraction of the requirements to input and output parameters allows for easier reasoning about the invariants and it is employed in subsequent sections.

However, an issue is how to identify which parameters are input and which output.
A straightforward automatic approach for distinguishing input parameters from the output ones is an iteration starting from the simple invariants and taking into account types of parameters already distinguished from the previous iterations.
The approach starts with the process invariants having only single parameter.
Such a parameter must be an \textit{output} one (otherwise the invariant would not produce any knowledge and the above mentioned IRM-SA idea would not hold).
Examples of such invariants are requirements 1(a) and 1(b).
Next, if a single attribute is present in multiple invariants, it can be assumed that it serves as an \textit{input} parameter.
Nevertheless, this is only an assumption and thus a fully automated approach is hard to achieve.
A possible solution is to use an assisted iterative approach, in which a tool identifies input and output parameters and the human designer confirms/reverts the decisions.

In some cases, computation associated with an abstract invariant may not have all the parameters precisely specified, as the particular parameters are unknown yet.
They may be specified in the child invariants and from the view of the higher-level invariant, they can be seen as an implementation detail.
In such cases, we use the notation \verb|V::?| to mark that the component \verb|V| participates in the invariant, but the specific attribute is specified later in a child invariant.
The final assignment of the parameter is up to the designer.

\subsection{Invariant refinement and composition}

In EBCS, communication between components is implicit via their knowledge sharing, which is conveyed via ensembles. 
An ensemble is thus specified via a condition determining when components are part of the particular ensemble and via knowledge that has to be interchanged.

Let us again assume the requirement 1(d) with the parameter abstraction

\begin{center}
\texttt{V::energy, V::POI, P::availability -> V::plan}.
\end{center}
As the parameters come from different components (\texttt{V} and \texttt{P}) but the computation can be performed only in a single component, it is clear that the invariant has to be \textit{refined} as a \textit{composition} of several invariants, from which at least one is an \textit{exchange} invariant (in the implementation, the exchange invariants results in the ensemble definition).
Such situations can be rather easily detected automatically based on the parameters' owners.

Nevertheless, refinement and composition of the \textit{abstract} invariants is more difficult as they represent high-level goals and can intentionally abstract from some ``implementation details'', i.e., omit some attributes.
For example, in Figure~\ref{fig:irm-e-cars-model}, composition of the invariants 5 and 6 means that the trip plan is computed first without any knowledge about availability of the parking places (the output parameter \texttt{V::planFeasibility} in the invariant 5) and then it is made more specific with information about the availability (the invariant 6).
Different composition of the lower-level invariants would lead to a completely different behavior.

Another issue in the automatic composition of invariants is a reasoning about \textit{situation-specific} requirements, which have to be grouped together according to a requirement they belong to.
The grouping is performed based on their output parameters and can result in duplication of invariant subtrees. 
However, identification of the right subtree to be duplicated is not straightforward.
Both these issues are very hard to solve automatically and we plan to further investigate possibilities of their automation in more detail.

\subsection{Model validation}

With the abstraction of invariants described in \ref{sec:knowledge_flow_recognition}, the IRM-SA model can be automatically validated according to the knowledge flow.
In particular, following checks can be performed:
\begin{itemize}
\item Configurations with missing input parameters can be discovered (i.e., an invariant producing the particular attribute is not included in the configuration due to a missing dependency relation).
\item Configurations with multiple invariants writing to the same attribute can be detected. 
\item Also, detection of unused output parameters or unused attributes can be performed.
\end{itemize}
All of these checks may point to flaws in the model and/or specification and discover them early.

\section{Related work}
\label{sec:related}

As far as we know, there are no attempts to automatize requirements processing for EBCS design.
Nevertheless, a related approach is described in~\cite{Casagrande2014}, in which authors propose an approach called NPL-KAOS that can automatically obtain a KAOS model from large volume of literature (KAOS~\cite{vanLamsweerde2008} is a goal-oriented requirement engineering method and it was one of inspirations for the IRM-SA method).
With the use of natural language processing tools and text mining techniques they process abstracts of scientific publications.
First, they detect goal-oriented keywords and then they use the Stanford parser to tag semantic structures.
From obtained semantic trees they extract goals and finally organize them into taxonomies.
The taxonomies are used to define relations between goals and this way they simulate the process of refinement.
Similarly to our approach, the authors try to automatically derive a model from textual data which would serve for purposes of requirements engineering.
However, their problem is different.
Their main goal is to help requirements engineer during the early stages of goal elicitation by extraction of main concepts from the large body of research abstracts.
Although the extraction process may miss some goals in the single abstract, with large number of abstracts they can count on the fact that the goal will be at the end noticed.
Contrary, our method is intended to process a single specification and therefore cannot reckon on this effect.

In \cite{DrechslerSW12}, the authors (semi-)automatically derive a UML model and OCL constraints from a specification and test cases, which are both written in natural language.
They employ formal methods to verify correctness of the derived design.
First, grammatical analysis is used to derive UML class diagrams from the test cases.
Then, the behavior of test cases is inspected and the UML sequence diagrams are derived.
In the next step, OCL constraints are deduced from the requirements and test cases.
Finally, verification of static aspects (UML class diagrams and OCL invariants) and dynamic aspects (satisfaction of specified method pre-conditions and post-conditions) is checked.
As in our approach, authors use the Stanford parser to get dependencies from the sentences.
They also employ WordNet (in their case, to distinguish components of the system and actors).
The main difference is that we directly target sCPS design and EBCS and therefore include an identification of \textit{process} and \textit{exchange} invariants, adaptability, etc.

\section{Conclusion}
\label{sec:conclusion}


In the paper, we have analyzed the IRM-SA  design method with respect to its possible automation with the help of natural language processing methods and tools.
We have identified steps that can be automated and sketched solutions.
As the automated natural language understanding is generally still a challenging issue, the full automation is hard (and in many cases impossible) to achieve.
Thus, we target a semi-automated system that guides the human designer, recommends solutions and validates the designers actions.

Currently, we plan to implement all the identified proposed solutions, to integrate them to the existing IRM-SA editor and to validate the resulting system on a real-life case-study.

Even though the proposed approaches are tailored to IRM-SA (which is currently tied with the DEECo component model), they can be reused in different contexts.
The IRM-SA method itself can be without changes applied to another ensemble-based component model (e.g., Helena~\cite{Hennicker_Helena_2014}) and the approaches proposed in this paper can be applied in tools for the KAOS method or similar ones.

\section{Acknowledgement}

This work was partially supported by the project no. LD15051 from COST CZ (LD) programme by the Ministry of Education, Youth and Sports of the Czech Republic and partially supported by Charles University institutional funding SVV-2016-260331.

\bibliographystyle{eptcs}
\bibliography{paper}

\end{document}